# Anomalously Robust Valley Polarization and Valley Coherence in Bilayer WS$_2$

Bairen Zhu[1+], Hualing Zeng[1,2+], Junfeng Dai[3], Zhirui Gong[1], Xiaodong Cui[1*]

1. Physics Department, University of Hong Kong, Hong Kong, China
2. Physics Department, Chinese University of Hong Kong, Hong Kong, China
3. Physics Department, South University of Science and Tech of China, Shenzhen, China

[+] Authors contribute equally

Coherence is a crucial requirement to realize quantum manipulation through light-matter interactions. Here we report the observation of anomalously robust valley polarization and valley coherence in bilayer WS$_2$. The polarization of the photoluminescence from bilayer WS$_2$ inherits that of the excitation source with both circularly and linearly polarized and retains even at room temperature. The near unity circular polarization of the luminescence reveals the coupling of spin, layer and valley degree of freedom in bilayer system, while the linear polarized photoluminescence manifests quantum coherence between the two inequivalent band extrema in momentum space, namely, the valley quantum coherence in atomically thin bilayer WS$_2$. This observation opens new perspectives for quantum manipulation in atomically thin semiconductors.

Tungsten sulfide WS$_2$ among the family of group VI transition metal dichalcogenides (TMDCs) is a layered compound with buckled hexagonal lattice. As WS$_2$ thins to atomically thin layers, WS$_2$ films undergo a transition from indirect gap in bulk form to direct gap at monolayer level with the band-edge located at energy-degenerate valleys (K,K') at the corners of the Brillouin zone.[1,2,3] Like the case in its sister compound monolayer $MoS_2$, the valley degree of freedom of monolayer $WS_2$ could be presumably addressed through nonzero but contrasting Berry curvatures and orbital magnetic moments which arise from the lack of spatial inversion symmetry at monolayers.[3,4] The valley polarization could be realized by control of the polarization of optical field through valley selective interband optical selection rules at K and K' valleys as illustrated in Figure 1a.[4-6] In monolayer $WS_2$ both the top of valence bands and the bottom of conduction bands are constructed primarily by the d-orbits of tungsten atoms which are remarkably shaped by spin-orbit coupling (SOC). The giant spin-orbit coupling splits the valence bands around the K (K') valley by 0.4eV while the conduction band is nearly spin degenerated.[7] As a result of time reversal symmetry the spin splitting has opposite signs at the K and K' valleys. Namely the Kramer's doublet $|K\uparrow\rangle$ and $|K'\downarrow\rangle$ are separated from the other doublet $|K'\uparrow\rangle$ and $|K\downarrow\rangle$ by the SOC splitting of 0.4eV. The spin and valley are strongly coupled at K(K') valleys and this coupling significantly suppresses spin and valley relaxations as both spin and valley index have to be changed simultaneously.

In bilayer $WS_2$, besides the spin and valley degrees of freedom, there exists an extra index: layer polarization which indicates the carriers' location, either up-layer or down-layer. Bilayer $WS_2$ follow a $2H$ packing order and the spatial inversion symmetry is recovered: each layer is 180 degree in plane rotation of the other with the tungsten atom of a given layer sitting exactly on top of the S atom of the other layer. The layer rotation symmetry switches K and K' valleys but leaves the spin unchanged, which results in a sign change for the spin-valley coupling from layer to layer (Figure 1b). From the simple spatial symmetry point of view, one might expect the valley dependent physics fades at bilayers owing to inversion symmetry, as the precedent of bilayer $MoS_2$.[8] Nevertheless the inversion symmetry becomes subtle if the coupling of spin, valley and layer indices is taken into account. Note that the spin-valley coupling in $WS_2$ is around 0.4eV (the counterpart in $MoS_2$ ~0.16eV) significantly higher than the interlayer hopping energy (~0.1eV), the interlayer coupling at K and K' valleys in $WS_2$ is greatly suppressed as indicated in Figure 1b.[7,9] Consequently bilayer $WS_2$ can be regarded as decoupled layers and it may inherit the valley physics demonstrated in monolayer TMDCs. In addition the interplay of spin, valley and layer degrees of freedom opens an unprecedented channel towards manipulations on quantum states.

Here we report a systemic study on the polarization resolved photoluminescence (PL) experiments on bilayer $WS_2$. The polarization of PL inherits that of excitations no matter circularly or linearly polarized, up to room temperature. The experiments demonstrate the valley polarization and valley coherence in bilayer $WS_2$ as a result of the coupling of spin, valley and layer degrees of freedom. Surprisingly the valley polarization and valley coherence in bilayer $WS_2$ are anomalously robust comparing with monolayer $WS_2$.

For comparison we first perform polarization resolved photoluminescence measurements on monolayer $WS_2$. Figure 2a shows the photoluminescence spectrum from monolayer $WS_2$ at 10K. The PL is dominated by the emission from band-edge excitons, so called "A" exciton at K and K' valleys. The excitons carry a clear circular dichroism under near resonant excitation (2.088eV) with circular polarization as a result of valley selective optical selection rules where the left-handed (right-handed) polarization corresponds to the interband optical transition at K(K') valley. The PL follows the helicity of the circularly polarized excitation optical field. To characterize the circular component in the luminescence spectra, we define a degree of polarization $P = \frac{I(\sigma+) - I(\sigma-)}{I(\sigma+) + I(\sigma-)}$ where $I(\sigma\pm)$ is the intensity of the right(left)-handed circular component. The luminescence spectra display a contrasting polarization for excitation with opposite helicities: $P=0.4$ under $\sigma+$ excitation and $P=-0.4$ under $\sigma-$ excitation on the most representative monolayer. For simplicity only the PL under $\sigma+$ excitation is shown in the plot. The degree of polarization P shows flat throughout the whole luminescence as shown in the inset of Figure 2a. These behaviors are fully expected in the mechanism of valley selective optical selection rules.[3,4] The degree of polarization decays with the lifted temperature and drops to 10% at room temperature (Figure 2b). The decay pattern cannot be interpreted by either off-resonance excitation or band edge phonon scattering.[10] And it decreases as the excitation energy shifts from the near resonant energy of 2.088eV to 2.331eV as illustrated in Figure 2c.

Next we turn to study the PL from bilayer $WS_2$. Figure 3 shows the PL spectrum from bilayer $WS_2$. The peak labeled as "I" denotes the interband optical transition from the indirect band gap and the peak 'A" corresponds to the exciton emission from direct band transition at K and K' valleys. Although bilayer $WS_2$ has an indirect gap, the direct interband optical transition at K and K' valleys dominates the integrated PL intensity as the prerequisite of phonon/defect scattering is waived for direct band emission and the direct gap is just slightly bigger than the indirect band gap in bilayers. Figure 3.a displays surprisingly robust PL circular dichroism of "A" exciton emission under circularly polarized excitations of 2.088eV (resonance) and 2.331eV (off-resonance). The degree of polarization of "A" exciton emission under near resonant σ± excitation is near unity (around 95%) at 10K and preserves around 60% at room temperature. In contrast, the emission originating from indirect band gap is unpolarized in all experimental conditions. It is unlikely that the high polarization in bilayers results from the isolation of the top layer from the environments as similar behaviors are observed in monolayer and bilayer $WS_2$ embedded in PMMA matrix or capped with a 20nm-thick $SiO_2$ deposition.

To exclude the potential cause of charge trapping or substrate charging effect, we study the polarization resolved PL of bilayer $WS_2$ with an electric back gate. Figure 4a elaborates the evolution of PL spectra under circularly polarized excitations of 2.088eV and a gate bias at 10K in a field effect transistor like device. The PL spectra dominated by "A" exciton emissions shows negligible change under the gate bias in the range of *-40V* to *20V*. The electric conductance measurements show the bilayer $WS_2$ stays at the electrically intrinsic state under the above bias range. The PL spectra can be safely recognized as emission from free excitons. As the gate bias switch to further positive side (>20V), the PL intensity decreases and emission from electron-bounded exciton "X⁻" so called trion emerges and gradually raise its weight in the PL spectrum.[11,12] The electron-exciton binding energy is found to be 45meV. Given only one trion peak in PL spectra, the interlayer trion (formed by exciton and electron/hole in different layers) and intralayer trion (exciton and electron/hole in the same layer) could not be distinguished due to the broad spectral width.[13] Both free exciton and trion show slightly red shifts with negative bias, presumably as a result of quantum confined stark effect.[14]

At all the bias conditions, the degree of polarization of free exciton and trion keeps unchanged within the experiment sensitivity as shown in Figure 4c. The bias independent degree of polarization rules out the possibility that effects of Coulomb screening, charge traps or charge transfers with substrates are the major causes for the robust circular dichroism in bilayers against monolayers.

One potential cause may result from the shorter lifetime of excitons at K and K' valleys for bilayer system. The band gap shifts from K and K' points of the Brillouin zone in monolayers to the indirect gap between the top of valence band at $\Gamma$ points and the bottom of conduction band in mid of K and $\Gamma$ points in bilayers. Our time resolved pump-probe reflectance experiments show that the exciton lifetime at K and K' valleys in bilayers is around tens ps, a fraction of that at monolayers (supplementary). If we assume the valley lifetime is the same for both monolayers and bilayers and the PL polarization $p = \dfrac{P_0}{1 + \tau/\tau_k}$ where $P_0$ is the theoretically limit of PL polarization, $\tau_k$ and $\tau$ denote the valley lifetime and exciton lifetime respectively, the shorter exciton lifetime will lead to significant higher PL polarization. If we assume that the $K \leftrightarrow K'$ intervalley scattering is at the same order in bilayers and monolayers, however, the difference in exciton lifetime between bilayers and monolayers is not overwhelming enough to be the major cause of robust polarization observed in the time integrated PL in bilayers nevertheless.

In monolayer WS$_2$ under circularly polarized resonant excitations, the depolarization mainly comes from the $K \leftrightarrow K'$ intervalley scattering. In bilayers, the depolarization could be either via $K \leftrightarrow K'$ intervalley scattering within the layer in a similar way as in monolayers, or via inter-layer hopping which also requires spin flip. As we discussed above, the interlayer hopping at K valley is suppressed in WS$_2$ as a result of strong SOC in WS$_2$ and spin-layer-valley coupling, which were experimentally proved by the circular dichroism in PL from bilayers. The robust polarization in bilayers implies that the intervalley scattering within layer is diminished compared with that in monolayers. There are two prerequisites towards intervalley scattering within layers: conservation of crystal momentum and spin flip of holes. The crystal momentum conservation could be satisfied with the involvement of phonons at K points in the Brillouin zone or atomic size defects, presumably sharing the similar strength in monolayers and bilayers. Spin-flip process could be realized by three different spin scattering mechanisms, namely D'yakonov-Perel (DP) mechanisms[15], Elliot-Yaffet (EY),[16] and Bir-Aronov-Pikus (BAP).[17,18] DP mechanism acts through a Lamor precession driven by electron wavevector $k$ dependent spin-orbit coupling. It is thought be negligible for spin flip along out-plane direction as the mirror symmetry with respect to the plane of W atoms secures a zero out-plane crystal electric field. Another possible driving force under DP mechanism could be the asymmetry owing to the interface with the substrate. This was excluded by the similar behaviors where the monolayers and bilayers $WS_2$ are embedded in PMMA matrix or capped with a thin layer of $SiO_2$. The negligible effect of electric gating on polarization also implies that DP mechanism is weak in mono- and bilayer $WS_2$; EY mechanism originates from scattering with phonons and defects. Its strength in bilayers and monolayers is likely to be at similar scale, and bilayers even have more low-frequency collective vibrational modes.[19] Therefore EY mechanism is unlikely to be the cause here; BAP mechanism originates from the electron-hole exchange interaction. In monolayer and bilayer TMDCs the optical features are dominated by the Wannier type yet giant excitonic effect and the exciton binding energy in such intrinsic 2D semiconductors is estimated to be *0.6~1eV.*[20,21] This giant exciton binding energy indicate a mixture of electron and hole wavefunctions and consequently strong exchange interaction which may contribute to the spin flip and intervalley scattering.[5,22] As the conduction band has a band mixing at K points, the spin flip of electron would be a quick process. An analogous scenario is that the spin of holes relaxes in hundreds of femtoseconds or less in GaAs as a result of band mixing and spin-orbit coupling. The electron spin flip could lead to hole spin flip via strong exchange interaction accompanying intervalley scattering which is realized by the virtual annihilation of a bright exciton in the K valley and then generation in the K' valley or vice versa.[22] This non single-particle spin relaxation leads to valley depolarization instead of the decrease of luminescence intensity which results from coupling with dark excitons. Generally the exciton binding energy decreases with the relaxation of spatial confinement. Whereas first principle calculation shows that monolayer and bilayer $WS_2$ shares the similar band dispersion and effective masses around K valley in their Brillouin zone as a result of spin-valley coupling.[7] It implies that the binding energy of excitons around K valley in bilayer $WS_2$ is similar to or slightly less than that in monolayer $WS_2$. As the exchange interaction is roughly proportional to the square of exciton binding energy, the spin flip rate and consequently intervalley scattering via exciton exchange interactions is presumably comparable or smaller to some extent in bilayer $WS_2$.(Supplementary) Nevertheless this is unlikely the major cause of the anomalously robust valley polarization in bilayer $WS_2$.

We also investigated the PL from bilayer $WS_2$ under a linearly polarized excitation. A linearly polarized light could be treated as a coherent superposition of two opposite-helicity circularly polarized lights with a certain phase difference. The phase determines the polarization direction. In semiconductors, a photon excites an electron-hole pair with the transfer of energy, momentum and phase information. The hot carriers energetically relax to the band edge in a quick process around $10^{-1} \sim 10^1$ ps through runs of inelastic and elastic scatterings, e.g. by acoustic phonons. During the quick relaxation process, generally the phase information randomizes and herein coherence fades. In monolayer TMDCs, the main channel for carrier relaxation is through intravalley scatterings including Coulomb interactions with electron (hole) and inelastic interactions with phonons, which are valley independent and preserve the relative phase between K and K' valleys.[23] In bilayer $WS_2$, the suppression of intervalley scattering consequently leads to the suppression of inhomogenous broadening in carrier's phase term. Subsequently the valley coherence demonstrated in monolayer WSe$_2$[23] is expected to be enhanced in bilayers.[13] The valley coherence in monolayer and bilayer WS$_2$ could be monitored by the polarization of PL under linearly polarized excitations.

Figure 5a shows the linearly polarized component of PL under a linearly polarized excitation of 2.088eV at 10K. The emission from indirect band gap is unpolarized and "A" exciton displays a pronounced linear polarization following the excitation. The degree of linear polarization $P = \dfrac{I(\parallel) - I(\perp)}{I(\parallel) + I(\perp)}$ is around 80%, where $I(\parallel)$ ($I(\perp)$) is the intensity of PL with parallel (perpendicular) polarization with respect to the excitation polarization. In contrast, the linear polarization is much weaker in monolayer samples (4% under the same experimental conditions, Figure 5b). As presented in Figure 5c, the polarization of "A" exciton is independent of crystal orientation and exactly follows the polarization of excitations. The degree of the linear polarization in bilayer WS$_2$ slightly decreases with the increased temperature and drops from 80% at 10K to 50% at room temperature (Figure 5d). This is the paradigm of the robust valley coherency in bilayer WS$_2$.

The linear polarization of both exciton and trion in bilayer, contrasting to the circular polarization which shows negligible dependence on the electric field in the range, shows a weak electric gating dependence as shown in Figure 5e. The PL linear polarization, presenting valley coherence, decreases as the Fermi level shifts to the conduction band. It doesn't directly effect in intervalley scattering within individual layers and make no observable change in circular dichroism. Nevertheless the electric field between the layers induces a layer polarization and slightly shifts the band alignments between the layers by different amounts in conduction and valence band,[13, 24] though the shift is indistinguishable in the present PL spectra due to the broad band width. The layer polarization and the shift of band alignments may induce a relative phase difference between two layers and therefore effects the PL linear polarization via interference. Further study is needed to fully understand the mechanism.

In summary we demonstrated anomalously robust valley polarization and valley polarization coherence in bilayer $WS_2$. The valley polarization and valley coherence in bilayer $WS_2$ is the direct consequence of giant spin-orbit coupling and spin valley coupling in $WS_2$. The depolarization and decoherence processes are greatly suppressed in bilayer, though the mechanism is ambiguous. The robust valley polarization and valley coherence make bilayer $WS_2$ an intriguing platform for spin and valley physics.

**Methods:**

Atomically thin $WS_2$ films were fabricated by mechanical exfoliation from a synthetic single crystal and the sample thickness was identified with optical microscope, photoluminescence spectroscopy and second harmonic generation.[7] All samples are mounted on heavily doped silicon substrates caped with 300nm-thick oxide. A polarization sensitive photoluminescence experiments was carried out with a confocal-like setup and the details could be found in Ref.[6] The carrier dynamics measurement was carried out using the time-resolved pump-probe technique and the details could be found in the supplementary information.

**Acknowledge:** The authors thank Prof. Wang Yao and Prof. Ming-Wei Wu for helpful discussion. H Zeng acknowledges the support of Direct Grant for Research 2013/14 (P/4053082) in CUHK. The work is supported by Area of excellency (AoE/P-04/08) and CRF of Hong Kong Research Grant Council.

**Author contributions**
BR Zhu, HL Zeng and XD Cui conceived and designed the experiments. BR Zhu, HL Zeng and JF Dai carried out the experiments and analysed data. XD Cui, HL Zeng and ZR Gong wrote the manuscript.

# Figure Caption

Figure 1. (a) Schematic of valley dependent optical selection rules and the Zeeman like spin splitting in the valence bands of monolayer $WS_2$. (b) Diagram of spin-layer-valley coupling in *2H* stacked bilayer $WS_2$. Inter-layer hopping is suppressed in bilayer $WS_2$ owing to the coupling of spin, valley and layer degree of freedom.

Figure 2. Photoluminescence of monolayer $WS_2$ under circularly polarized excitation. (a) Polarization resolved luminescence spectra with $\sigma+$ detect (red) and $\sigma-$ detect (black) under near-resonant $\sigma+$ excitation (2.088eV) at 10K. Peak A is the excitonic transition at band edges of K(K') valleys. Opposite helicity of PL is observed under $\sigma-$ excitation. Inset presents the circular polarization at peak position. (b) The circular polarization as a function of temperature. Black is data. The curve (red) is a fit following a Boltzman distribution where the intervalley scattering by phonons is assumed. (c) Photoluminescence spectrum under off-resonant $\sigma+$ excitation (*2.33eV*) at *10K*. The red (black) curve denotes the PL circular components of $\sigma+(\sigma-)$. A circular polarization P of *16%* is observed.

Figure 3. Photoluminescence of bilayer $WS_2$ under circularly polarized excitation.
(a) Polarization resolved luminescence spectra with components of $\sigma+$ (red) and $\sigma-$ (black) under near-resonant $\sigma_+$ excitation (2.088eV) at 10K. Peak A is the excitonic transition at band edges of direct gap. Peak I is the indirect band gap emission showing no polarization. Inset presents the circular polarization of the A excitonic transition at peak position. Opposite helixicity of PL is observed under $\sigma-$ excitation. (b) The circular polarization as a function of temperature (black). The curve (red) is a fit following a Boltzman distribution with where the intervalley scattering by phonons is assumed. (c) Photoluminescence spectrum of components of $\sigma+$ (red) and $\sigma-$ (black) under off-resonant $\sigma+$ excitation (2.33eV) at 10K. A circular polarization P of 70% is only observed for the transition (A exciton) at K(K') valleys.

Figure 4. Electric doping dependent photoluminescence spectrum of bilayer $WS_2$ FET. (a) Luminescence spectra of bilayer $WS_2$ at different gate voltage under near-resonant $\sigma+$ excitation (2.088eV) at 10K. X denotes neutral exciton and $X^-$ is trion. Green curve is line shape fitting with two Lorentzian peak fits (peak I and $X^-$) and one Gaussian peak fit (peak X). (b) Intensity of exciton and trion emissions versus gate. Upper panel is the gate dependent integral PL intensity of exciton (X) and trion ($X^-$) in red. Lower panel shows relative ratio of integral PL intensity of exciton: trion as a function of gate voltage in black. (c) Degree of circular polarization of exciton (X, red) and trion ($X^-$, blue) versus gate.

Figure 5. Linearly polarized excitations on monolayer and bilayer $WS_2$.
(a) Linear polarization resolved luminescence spectra of bilayer $WS_2$ under near-resonant linearly polarized excitation (2.088eV) at 10K. Red/black is the parallel/cross polarization detect with respect to the linear polarization of excitation source. A linear polarization of 80% is observed for exciton "A" while the indirect gap transition (I) is unpolarized. (b) Linear polarization resolved luminescence spectra of monolayer $WS_2$ under near-resonant linearly polarized excitation (2.088eV) at 10K. Red/black is the parallel/cross polarization detect with respect to the linear polarization of excitation source. The linear polarization for exciton "A" in monolayer $WS_2$ is much weaker with a maximum value of 4%. (c) Polar plot for intensity of the exciton "A" in bilayer $WS_2$ (black) as a function of the detection angle at 10K. Red curve is a fit following $\cos^2(\theta)$. (d) The linear polarization of exciton "A" in bilayer $WS_2$ (Black) as a function of temperature. The curve (red) is a fit following a Boltzman distribution where the intervalley scattering by phonons is assumed. (e) Electric doping dependence of the linear polarization of exciton "A" in bilayer $WS_2$ at 10K.

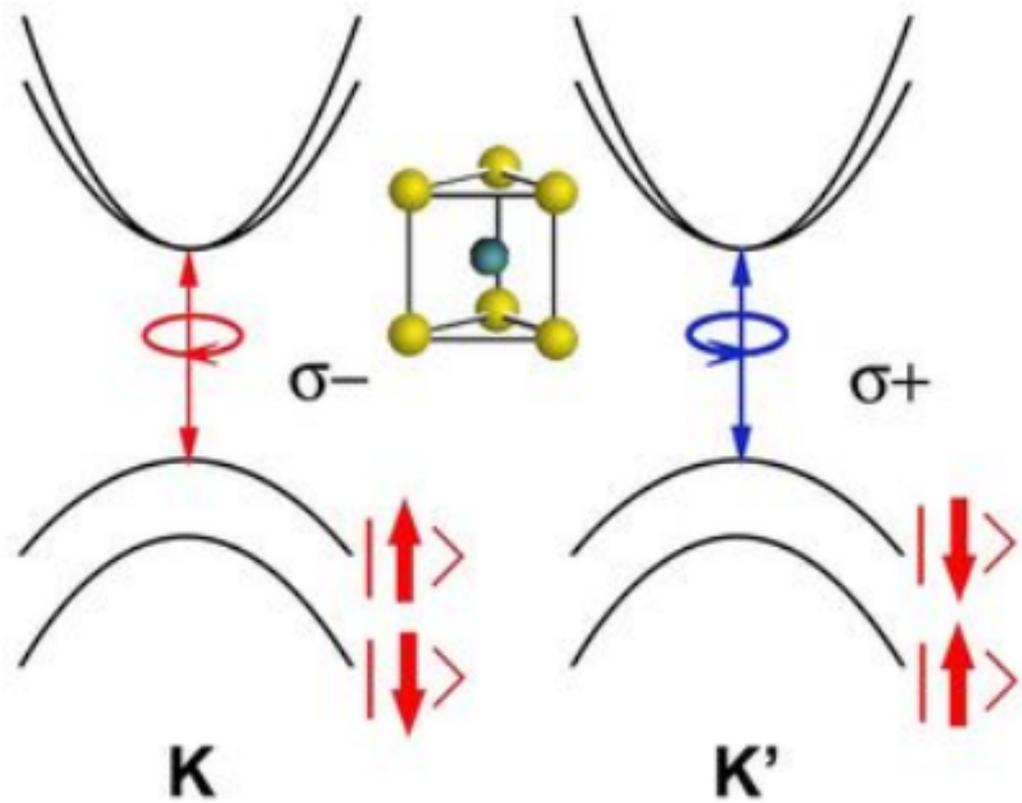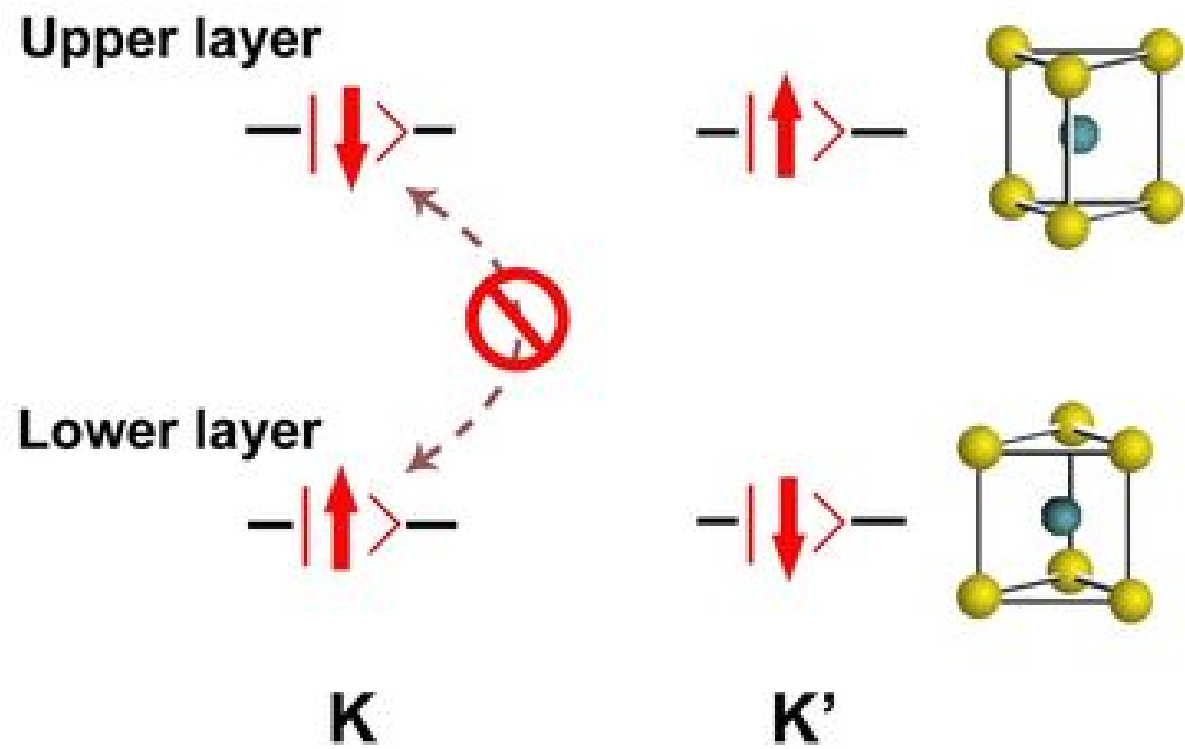

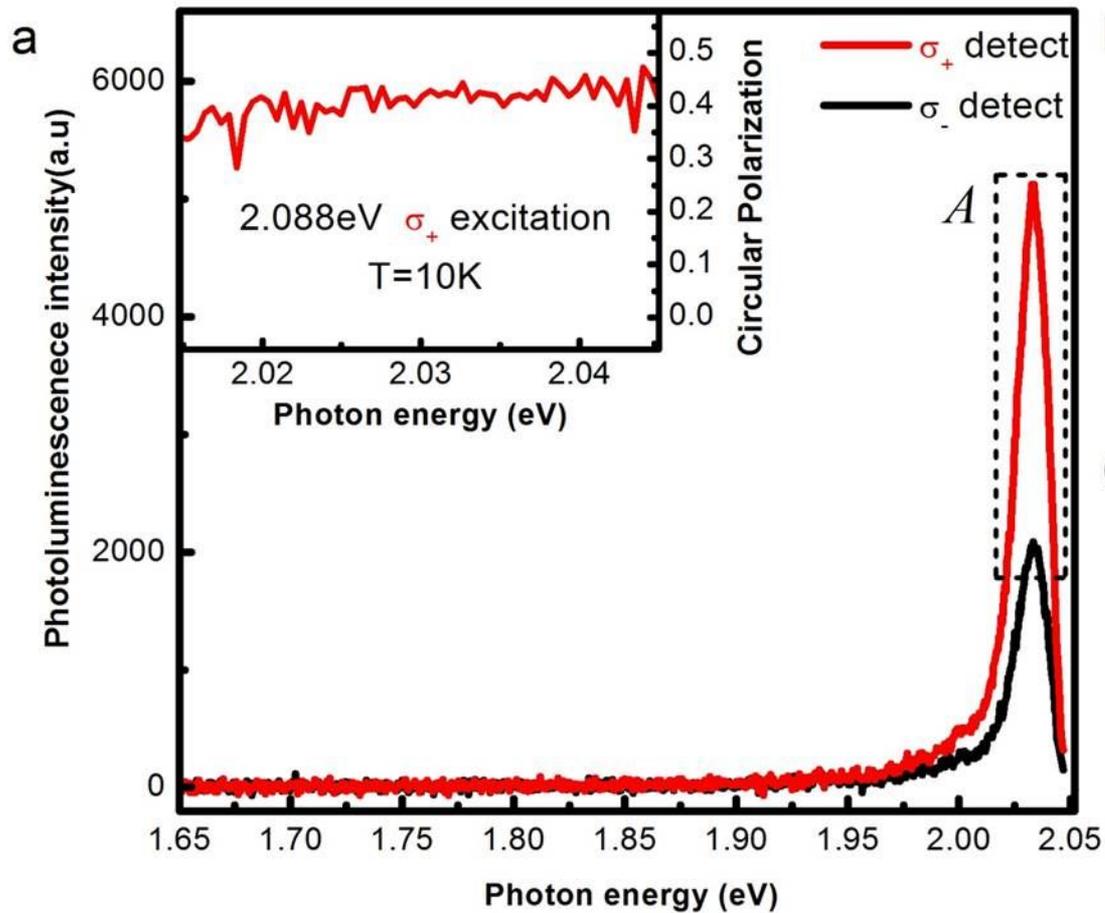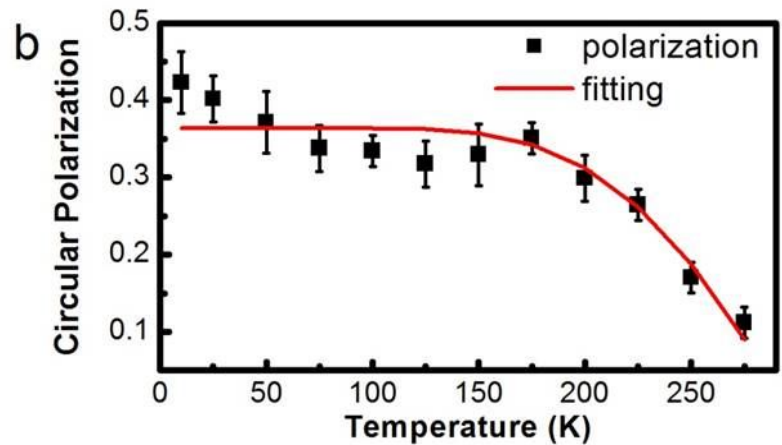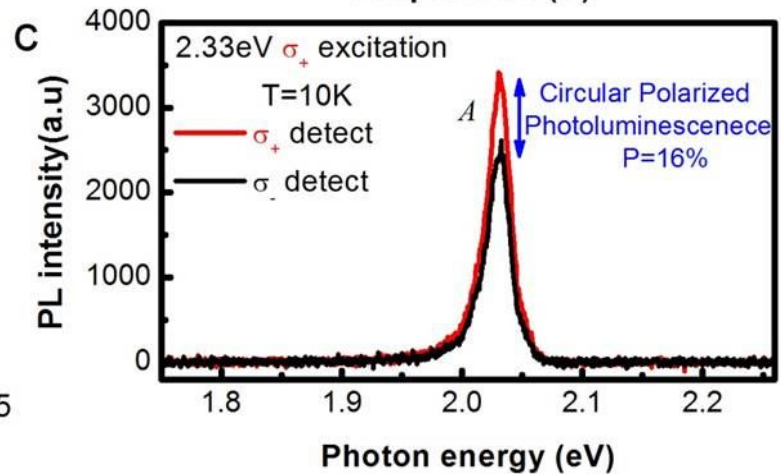

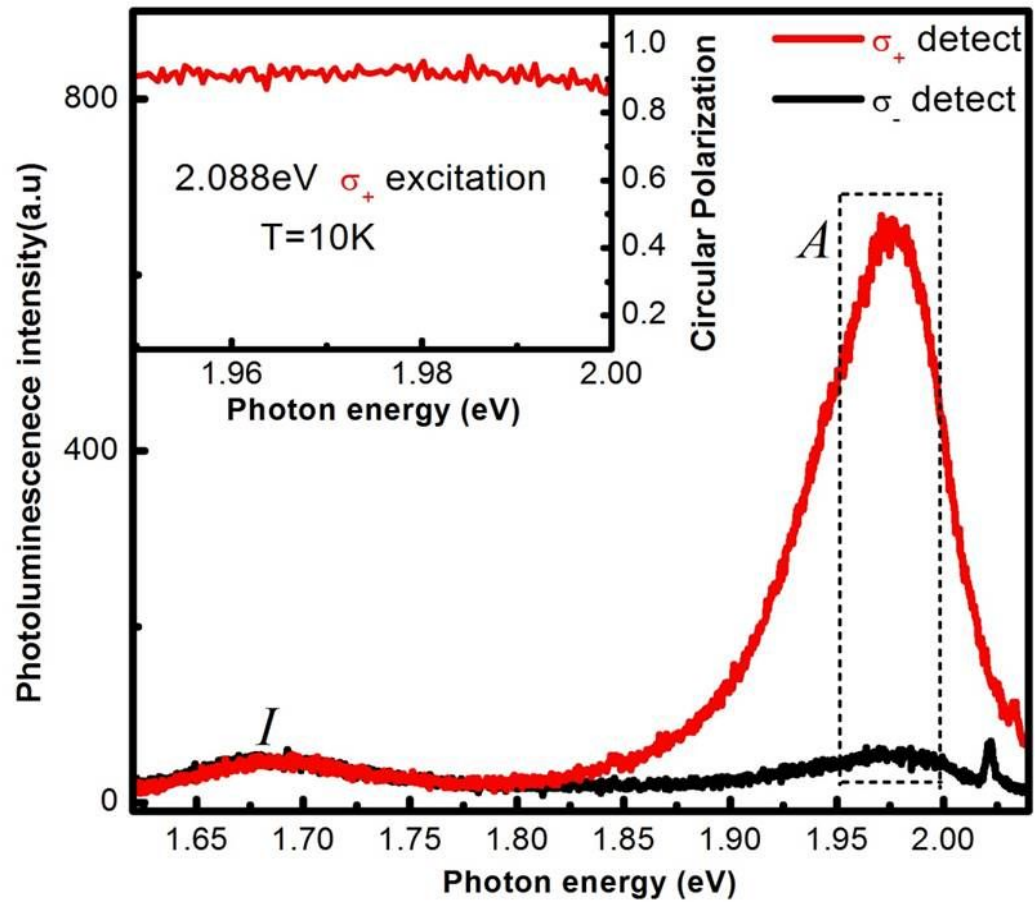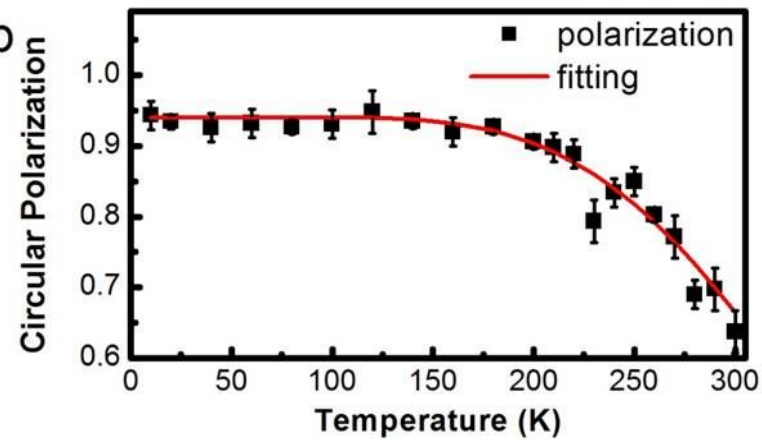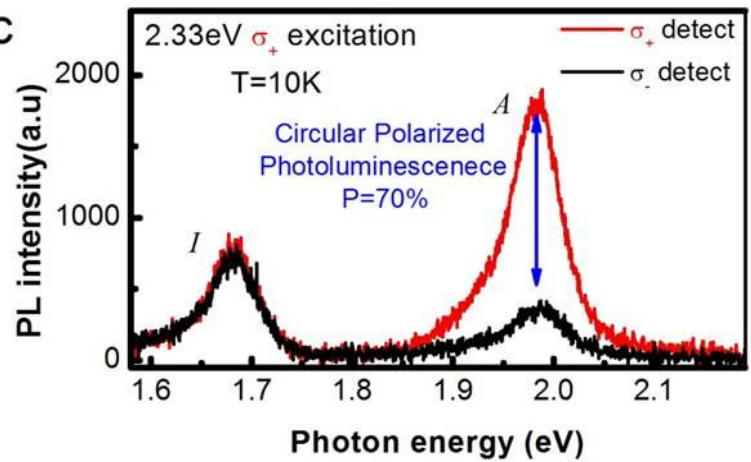

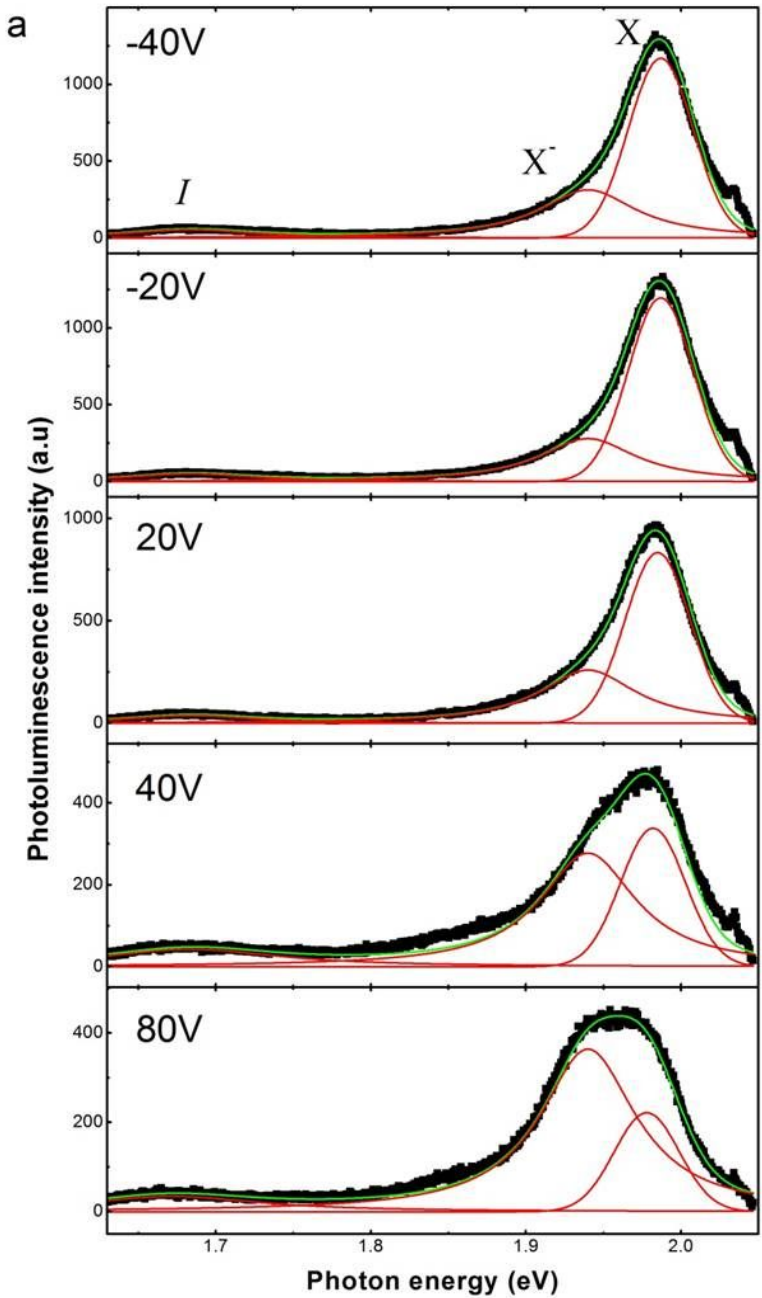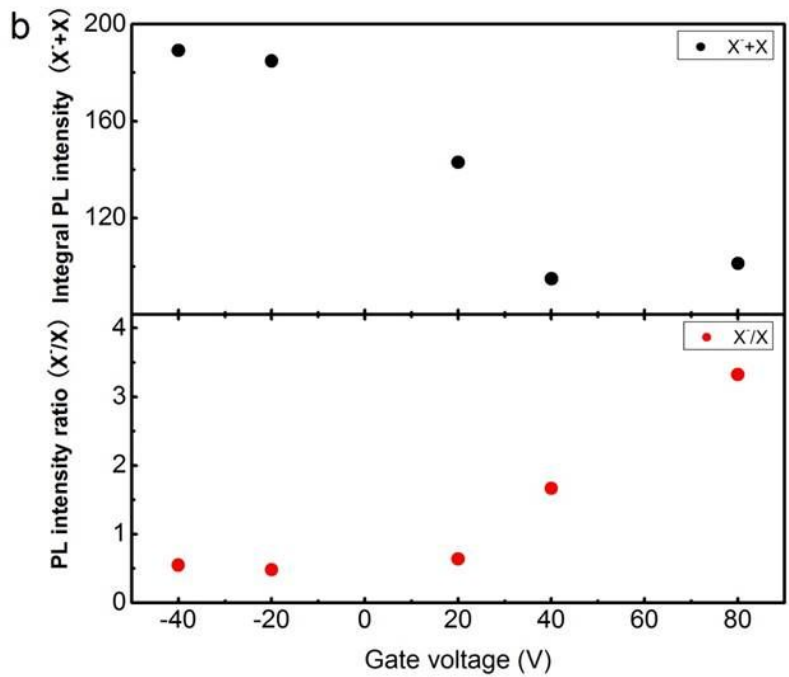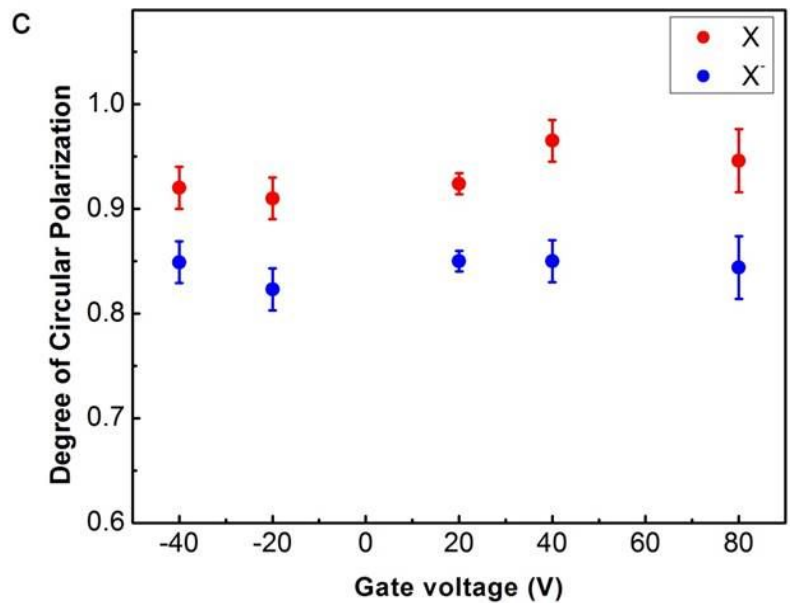

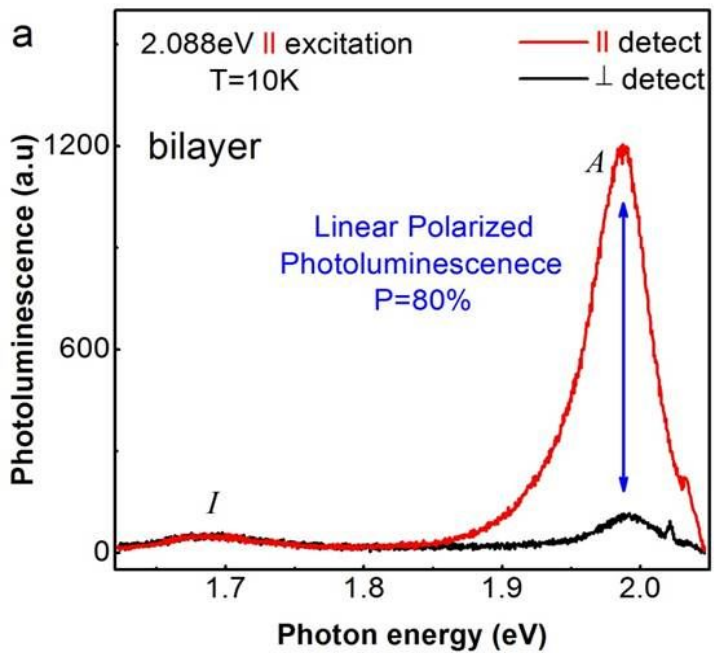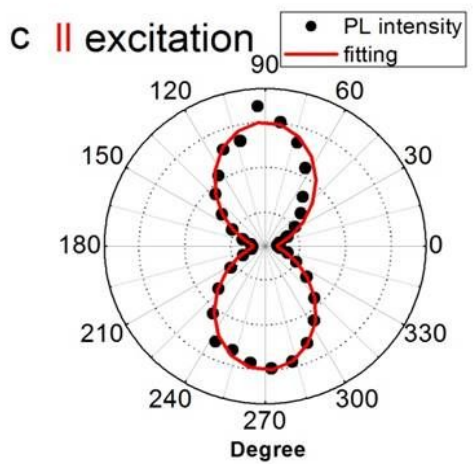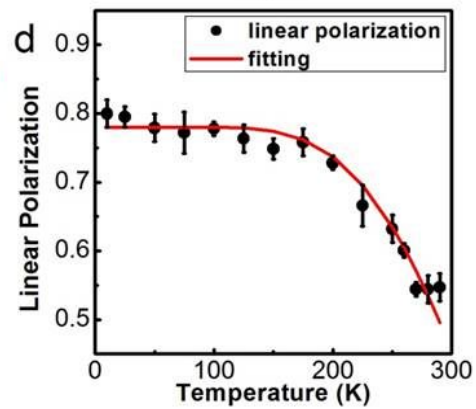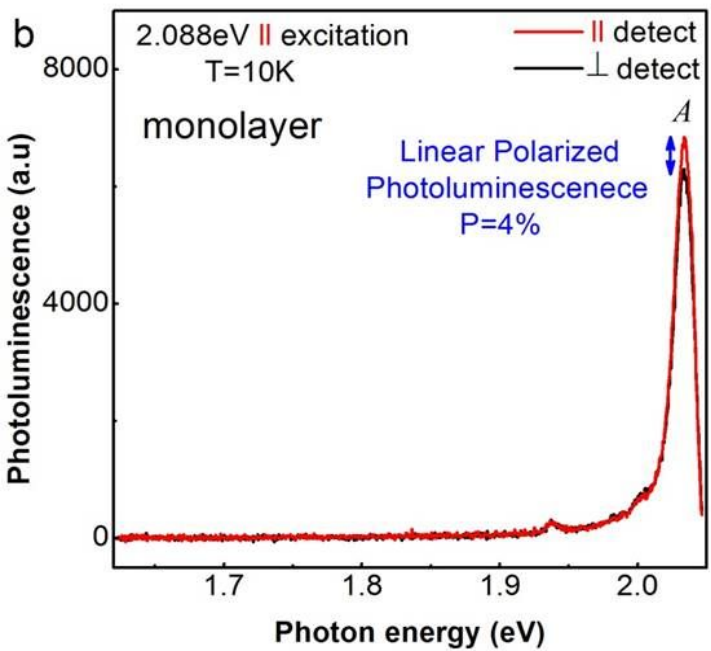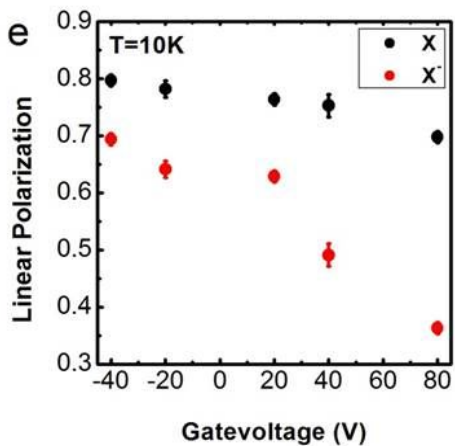